\documentclass[10pt,letterpaper]{article}
\usepackage[top=0.85in,left=2.75in,footskip=0.75in]{geometry}

\usepackage{amsmath,amssymb}

\usepackage{changepage}

\usepackage{textcomp,marvosym}

\usepackage{cite}

\usepackage{nameref,hyperref}

\usepackage[right]{lineno}

\usepackage[nopatch=eqnum]{microtype}
\DisableLigatures[f]{encoding = *, family = * }

\usepackage[table]{xcolor}

\usepackage{array}

\newcolumntype{+}{!{\vrule width 2pt}}

\newlength\savedwidth

\raggedright
\usepackage[aboveskip=1pt,labelfont=bf,labelsep=period,justification=raggedright,singlelinecheck=off]{caption}

\makeatletter
\renewcommand{\@biblabel}[1]{\quad#1.}
\makeatother

\usepackage{lastpage,fancyhdr,graphicx}
\usepackage{epstopdf}
\fancyheadoffset[L]{2.25in}
\fancyfootoffset[L]{2.25in}
\begin{document}
\vspace*{0.2in}

\begin{flushleft}
{\Large
\textbf{The cyborg fly: reflexes survive engineered actuators, coordination needs stand-ins}
}
\newline
\\
Mauricio Mu\~noz Arias\textsuperscript{1*}
\\
\bigskip
\textbf{1} Engineering and Technology Institute Groningen (ENTEG), University of Groningen, Groningen, The Netherlands
\\
\bigskip
* m.munoz.arias@rug.nl
\end{flushleft}

\section*{Abstract}
Two wiring diagrams now cover the whole central nervous system of an adult fruit fly: one female, one male. Can such a diagram, run as a network of spiking neurons, drive a body? And if we swap the body's muscles for engineered actuators, what survives? We ran both connectomes, BANC and MaleCNS, as leaky integrate-and-fire networks of about 160,000 neurons each, and closed the loop through simulated sensors, actuators and bodies for walking and flight. The actuators are port-Hamiltonian, so energy is audited at every step, and swapping one for another changes only a configuration file. Null networks shuffled at five levels, from the whole nervous system to synapse counts, test whether a result depends on the wiring. The wiring produced specific functions: sugar on the labellum drove the proboscis motor neuron, leg joints resisted imposed movement, leg circuits oscillated, a looming disc drove a take-off signal on its side (in MaleCNS), and descending neurons throttled and steered flight. The resistance reflex survived 20 of 20 actuator settings in MaleCNS and 19 of 20 in BANC, including a Hill-type muscle pair and a geared DC motor; the actuator set walking speed and energy cost instead. Coordinated six-leg walking, flight stability and the correction of perturbations did not emerge. Each needed a labelled engineered stand-in, which the connectome could command. Such stand-ins can also manufacture an apparent dependence on wiring: in a rebuild of a published closed loop, odour navigation came from a hand-written rule, and the network entered only through one valence number. The two animals differed on some functions, and the differences traced to wiring, such as a haltere drive of the flight power muscles found only in MaleCNS. A whole-CNS connectome, embodied, gives reflexes, rhythms and command signals, but not the coordination that turns them into behaviour.

\section*{Author summary}
A fruit fly's nervous system has been mapped down to every connection. Does the map alone make a fly? We built a simulated fly whose brain and nerve cord are exactly this map, run as spiking neurons, and whose senses and muscles are engineered. Then we asked what the wiring does on its own. It does a lot. It extends the proboscis to sugar, stiffens a leg that is pushed, flinches from a looming shadow, and sets the power and direction of flight. It does not walk on six legs, and it does not keep the fly upright in the air. For that we had to add engineered controllers, and the wiring could steer them. Such controllers can mislead: when we rebuilt another group's navigating fly, its success came from a hand-written rule, not from the map. We also replaced the muscles with motors and artificial muscles whose energy we track exactly. The reflexes survived; the speed and cost of walking changed. Two maps, from a female and a male fly, agreed on some behaviours and not on others, and we traced the differences to specific connections.

\section*{Introduction}
A wiring diagram is not a nervous system. It lists who connects to whom, and how often, but not how strongly, how fast or under what chemical conditions. Behaviour, moreover, does not stop at the edge of the nervous system. It runs through muscles, joints and the world, and comes back through the senses. To ask what wiring does, we must therefore put it in a body. And once it is in a body, a second question opens. The body is the part of the loop that engineering can replace. If the fly's muscles were motors, or artificial muscles, or springs, which of its behaviours would survive? This is the cyborg question, and it cuts two ways. It tests how much of a behaviour lives in the wiring rather than in the mechanics. And it asks whether a biological controller can run engineered hardware at all.

Wiring alone already says a surprising amount. A leaky integrate-and-fire model of the whole adult fly brain, with synapse counts as weights and transmitter identity as sign, predicted which neurons respond to taste and which drive feeding~\cite{shiu2024}. Connectome-constrained models of the visual system predicted motion tuning across cell types~\cite{lappalainen2024}, and rate models of the nerve cord found a central pattern generator for walking in four datasets~\cite{pugliese2026}. Two developments now let us close the loop. Two connectomes cover the whole central nervous system, brain and nerve cord together: BANC from a female~\cite{bates2026} and MaleCNS from a male~\cite{berg2026}. And physics models of the fly body exist for walking, NeuroMechFly~\cite{wangchen2024}, and for flight, flybody~\cite{vaxenburg2025}, both built on MuJoCo~\cite{todorov2012}.

Several groups have begun to join the two. A company demonstration drove NeuroMechFly from a brain-only connectome through a few hand-chosen descending neurons~\cite{eon2026}. A recent preprint ran BANC's whole nervous system in NeuroMechFly and reported odour navigation that rewired networks could not reproduce~\cite{minnick2026}. Others have run both connectomes in walking bodies with null networks graded by cell type~\cite{amandeep2026}, or catalogued which behaviours emerge and which do not~\cite{aldunate2026}. A critique has asked whether any of this needs the real wiring at all, since even a worm connectome can steer a fly body through a learned decoder~\cite{brunton2026}. In all of this work the body's actuators are fixed. To our knowledge, no study has held a whole-CNS connectome fixed while replacing the actuators it drives, audited the energy that flows between network and body, or flown a fly from motor neuron spikes alone. Without that, one cannot say whether a behaviour belongs to the wiring or to the mechanics that carry it out.

In this work we close the loop around both whole-CNS connectomes and swap the body's actuators. Our contributions are four.
(i) A port interface between network and body. Every sensor and actuator is a port with a neural side and a physical side, so an actuator can be swapped by a change of configuration without touching the network. The actuators are port-Hamiltonian~\cite{maschke1992,vanderschaft2014}, and their energy is audited at every step.
(ii) A closed loop around the whole central nervous system of two flies, female and male, with walking, flight, and senses from vision to taste. Every engineered stand-in is labelled and switchable, and every result is tested against null networks shuffled at five levels, including synapse counts and leg segment, and on both connectomes.
(iii) A test of which behaviours survive the actuator swap, with four actuator families, from a first-order lag to a Hill-type muscle and a geared motor, and with an asynchronous flight muscle driven only by motor neuron spikes.
(iv) A rebuild of the published odour-navigation loop~\cite{minnick2026}, to ask where its dependence on wiring comes from.

The answer is mixed, and we think it is clear. The wiring gives specific reflexes, rhythms and command signals, and the reflexes survive every actuator we tried. It does not give coordinated walking or stable flight. Those needed stand-ins, which the connectome could command but not replace. And stand-ins must be read with care: when we rebuilt the published odour-navigation loop, its dependence on wiring came from a hand-written rule fed by one number from the network.

The paper is organised as follows. Results first describe the closed loop, then what the wiring produces against null networks, how engineered actuators change walking, the senses, flight, the differences between the two connectomes, and the rebuilt odour-navigation loop. The Discussion asks what lives in the wiring, what is missing for coordination, and how stand-ins can mislead. Materials and methods give the neuron model, the port-Hamiltonian form of each actuator, the stand-ins, the null networks and the statistics.

\section*{Results}

\subsection*{A closed loop around two whole nervous systems}

\begin{figure}[!h]
\includegraphics[width=\textwidth,height=0.8\textheight,keepaspectratio]{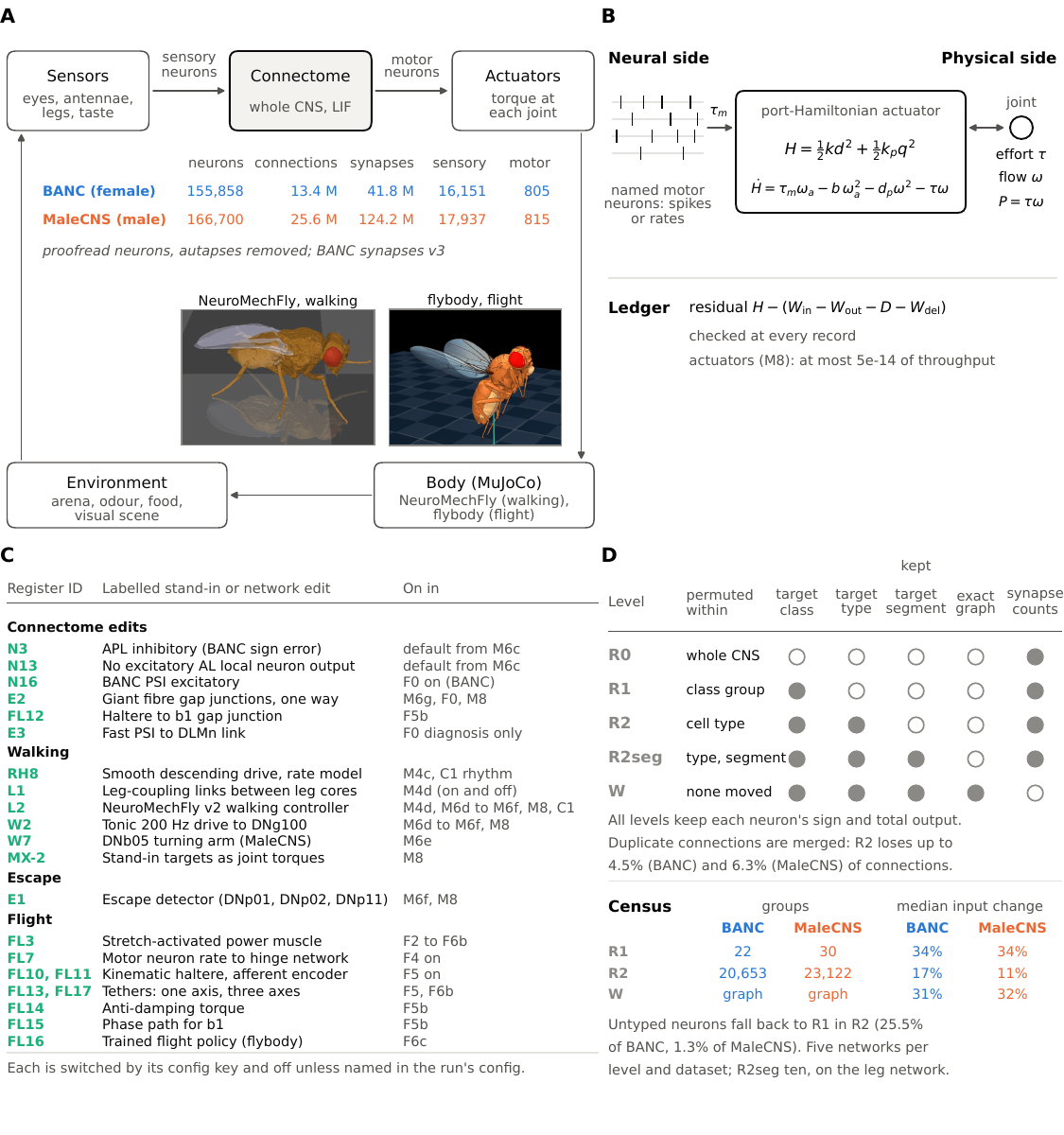}\par
\caption{\textbf{A closed loop around two whole-CNS connectomes.}
A: The loop, from environment to sensors, connectome, actuators and body. Counts are graph nodes (proofread neurons) and synapses with autapses removed; BANC synapse release v3. Stills: the two body models in MuJoCo, NeuroMechFly walking under the labelled stand-in (0.9~s) and flybody hovering under its trained policy with BANC commands (0.91~s). B: A port has a neural side (named motor neurons and their spikes) and a physical side (an effort and a flow); actuator ports are port-Hamiltonian, with storage $H$, dissipation and an energy ledger. Symbols as in Materials and methods. C: The register of labelled stand-ins and network edits, with register IDs and the experiments in which each is on. D: Null networks at five levels. Dots show what each level keeps; numbers give the census of groups and the median change in a neuron's total input (BANC v888 v3, MaleCNS v1.0). Drawn by a script written with Claude Code (see Use of AI tools).}
\label{fig1}
\end{figure}

We built the loop in the order a signal travels: environment, sensors, sensory neurons, connectome, motor neurons, actuators, body, and back to the environment (Fig~\ref{fig1}A). The nervous system is one of two connectomes. BANC, from a female, gave 155,858 neurons and 41.8 million synapses~\cite{bates2026}. MaleCNS, from a male, gave 166,700 neurons and 124 million synapses~\cite{berg2026}. Both counts match the published ones within 0.05\%. We used them as published, with labelled edits: by default one sign fix and the removal of one antennal-lobe pathway, and in BANC a second sign fix for escape and flight (Materials and methods).

Each neuron is a leaky integrate-and-fire unit with the parameters of Shiu et al.~\cite{shiu2024}. A synapse count times a fixed weight sets the strength of each connection; the predicted transmitter of the presynaptic neuron sets its sign. Our solver reproduced Brian2 and an independent implementation spike for spike. On the published FlyWire brain it gave the published response of the proboscis motor neuron MN9 to labellar sugar, 66.5 against 67.0~Hz at 100~Hz input and 91.3 against 93.3~Hz at 200~Hz. On BANC the response was graded but about half as large, and the whole nervous system answered within 2~Hz of the brain alone at every input rate. We chose the weight per synapse as the largest value that kept activity selective: 0.275~mV for BANC, 0.18~mV for MaleCNS.

Between the network and the body sit ports (Fig~\ref{fig1}B). A sensor port turns a physical quantity into firing rates of named sensory neurons: joint angle and load for leg proprioceptors, contact for taste, odour concentration for olfactory receptor neurons, antennal rotation for Johnston's organ, images for the eyes. An actuator port turns motor neuron spikes into torque at a joint. Its physical side is an effort and a flow, here torque and angular velocity, so the power it delivers is always defined. Actuators are port-Hamiltonian: each declares the energy it stores, the energy it dissipates and the power at its ports, and a ledger checks the balance at every recorded step.

Some behaviours did not come from the wiring. Where a later experiment needed them, we added an engineered stand-in and labelled it: a walking controller commanded by descending neurons, a detector that reads escape neurons, one-way gap junctions on the giant fibre, and a trained flight policy whose turning is commanded by a descending neuron (Fig~\ref{fig1}C). Every stand-in can be switched off, and every result names the stand-ins it used. A register of about 150 numbered assumptions records each choice, its source and how it could be tested (S1 Table).

To ask whether a result depends on the wiring, we built null networks at five levels (Fig~\ref{fig1}D). R0 shuffles every target across the nervous system, keeping each connection's source and synapse count. R1 shuffles targets within a class, such as left nerve cord interneurons. R2 shuffles within a cell type; untyped neurons, a quarter of BANC, fall back to R1. R2seg also keeps the leg segment, and we used it on the leg network. W keeps the graph exactly and shuffles synapse counts among each neuron's outputs. A result depends on the wiring at a given level if the real network falls outside the range of all null networks at that level. With five networks this happens by chance one time in three, so we treat a single such result as an indication and look for agreement across levels, readouts and connectomes.

\subsection*{What the wiring produces, and what depends on it}

\begin{figure}[!h]
\includegraphics[width=\textwidth,height=0.8\textheight,keepaspectratio]{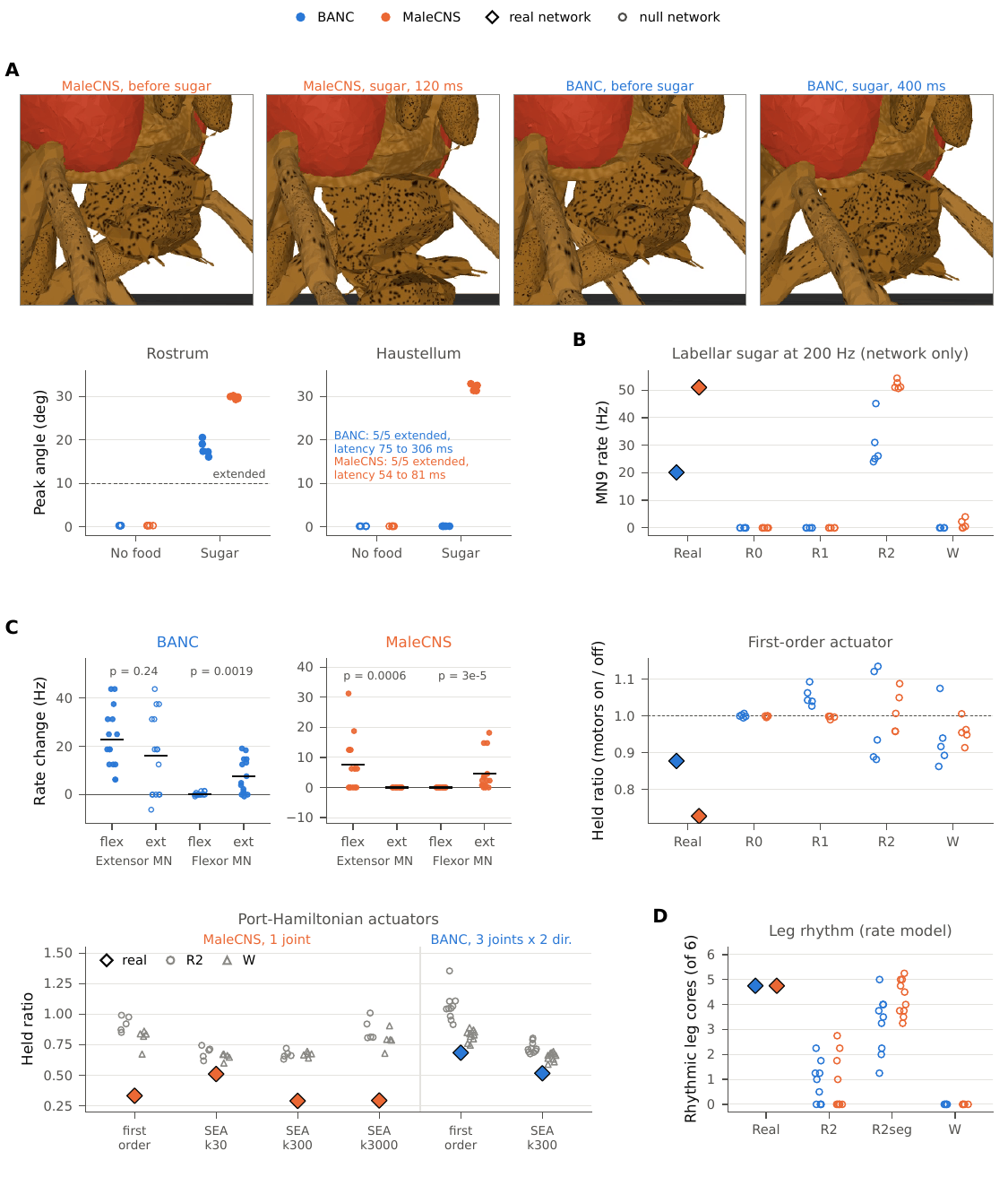}\par
\caption{\textbf{Functions the wiring produces, against null networks.}
A: Proboscis extension to labellar sugar. Top: video stills of the head from the side, before sugar (192~ms) and after it (sugar from 200~ms): MaleCNS at 320~ms, 120~ms after onset, with rostrum and haustellum extended; BANC at 600~ms, 400~ms after onset, with only the rostrum lowered. One video run per connectome, same settings as the counted runs; the drop is not drawn. Below: peak rostrum and haustellum angle per body run (5 per condition), no food against sugar. Dashed line, the 10-degree extension threshold. B: MN9 rate at 200~Hz sugar input, network only. Each point is one network (mean of 4 seeds); 5 networks per family. C: Resistance reflex. Left: motor neuron responses to imposed flexion and extension with motors off, 15 runs per direction pooled over 5 legs; filled points resist, bars are means, $p$ from Mann-Whitney tests. Middle: held ratio with the first-order actuator, real against 5 networks per family. Bottom: port-Hamiltonian actuators against R2 (circles) and W (triangles); MaleCNS one push (left middle femur-tibia flexion), 5 + 5 networks; BANC three joints and both directions, 10 + 10 networks. SEA $k$, $b$: series-elastic stiffness and damping. D: Leg rhythm: oscillating leg cores for the real network, R2 and R2seg (10 networks each) and W (5), 4 seeds each, at command level 800 (BANC) and 400 (MaleCNS). R2seg keeps cell type, side and neuromere. Drawn by a script written with Claude Code (see Use of AI tools).}
\label{fig2}
\end{figure}

Three functions came from the wiring with no engineered help (Fig~\ref{fig2}).

\textbf{Feeding.} Sugar on the labellum extended the proboscis in five of five runs in each connectome. MaleCNS also extended the distal haustellum, which BANC never did. Latency was 54 to 81~ms in MaleCNS and 75 to 306~ms in BANC. Under null networks, MN9 fell silent in R0 and R1, and nearly so under W (BANC 0~Hz, MaleCNS 0 to 4~Hz at 200~Hz input). It kept its rate in every R2 network: BANC 20.1~Hz real against 24 to 45~Hz, MaleCNS 51.0 against 50.6 to 54.4~Hz. Feeding needs cell types wired to cell types; within a type, the exact partners do not matter. The synapse counts do.

\textbf{A resistance reflex.} When we pushed a leg joint, leg proprioceptors drove the motor neurons that resist the push. In MaleCNS both arms of the reflex were significant: extensor motor neurons rose by 7.5~Hz when the joint was flexed and flexor motor neurons by 4.6~Hz when it was extended ($p=0.0006$ and $p=0.00003$, 15 runs per direction). BANC showed only the flexor arm. With the actuators on, the joint moved less late in the push than with the motors off: a held ratio of 0.73 in MaleCNS and 0.88 in BANC. Swapping the proprioceptor tuning reversed the sign of the effect. With port-Hamiltonian actuators the reflex lay outside every R2 and W null network. In MaleCNS, for one push and four actuators, the held ratio was 0.29 to 0.51 against 0.60 to 1.01 for ten null networks each. In BANC, tested over three joints and both directions, it lay outside all 20 for both actuators tested. In BANC with the first-order actuator, by contrast, it lay inside the null range. The reflex is wiring-specific below the level of cell type wherever the test could separate it.

\textbf{A leg rhythm.} Spiking leg circuits did not oscillate: at most 5\% of 4,320 runs showed a rhythm. Run as a rate model with a smooth descending command, following Pugliese et al.~\cite{pugliese2026}, the leg circuits did oscillate. At the settings of the rhythm study, MaleCNS oscillated in 5.25 of six leg cores on average and BANC in 3 of 6, at 11 to 14~Hz; BANC's right hind leg was always silent. For the null test we used one command level per connectome, at which the real network oscillated in 4.75 cores in both (Fig~\ref{fig2}D). Under R2 it fell to 0 to 2.75 cores, and under R2seg, which keeps each neuron in its own leg segment, it came back to 1.25 to 5.25. Plain R2 had moved inputs between legs. Under W the rhythm was lost (at most 0.25 cores). At other command levels R2seg spread more widely, and in MaleCNS at a low level it did no better than R2. The rhythm is a property of cell types once the segment is kept, with this caveat.

What the wiring did not produce matters as much. The connectome never held the body upright: standing height stayed at the limp value, 0.75 to 0.86~mm against 1.1 to 1.5~mm for replayed standing. Six legs on the ground stepped 1.5 legs at a time on average and did not walk; the fly tipped over (S1 Video). The front legs alternated in antiphase in 88 to 100\% of runs when sensory feedback was on, at 8.5 to 9.5~Hz against about 11~Hz in flies, but the six legs never coordinated.

\subsection*{Engineered actuators keep the reflex and change the gait}

\begin{figure}[!h]
\includegraphics[width=\textwidth,height=0.8\textheight,keepaspectratio]{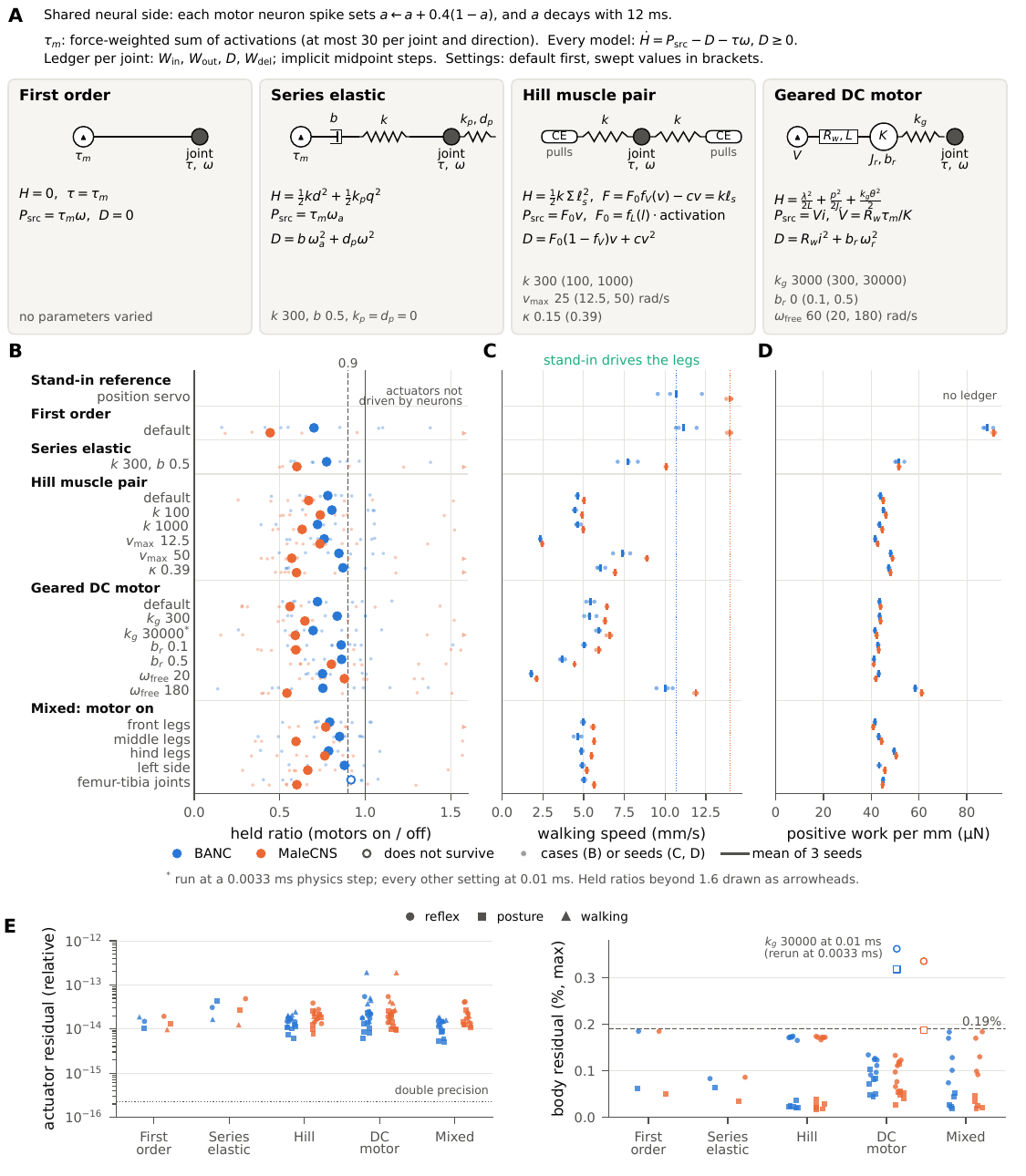}\par
\caption{\textbf{Engineered actuators keep the reflex and change the gait.}
A: The four actuator models and their port-Hamiltonian structure; symbols as in Materials and methods. B: Held ratio for each of 20 actuator settings, both connectomes. Large markers are medians over 8 leg and direction cases; small dots are cases (3 replicates each, 10 for settings at or above 0.8). Open circles do not survive; dashed line at 0.9. Six case ratios above 1.6 are drawn as arrowheads at the edge. C: Walking speed under the stand-in for each actuator; dots are 3 seeds, bars their mean; dotted lines, the position-controlled stand-in. D: Positive work per millimetre walked, same runs (µN). E: Energy balance: largest actuator residual relative to throughput per setting (reflex, posture, walking), and largest body residual (reflex, posture); dashed line 0.19\%. Open markers: gear stiffness 30,000 at 0.01~ms; that setting is shown at 0.0033~ms elsewhere, all others at 0.01~ms. Drawn by a script written with Claude Code (see Use of AI tools).}
\label{fig3}
\end{figure}

We then kept the connectome fixed and replaced the actuators (Fig~\ref{fig3}). Four kinds, all port-Hamiltonian, carried the same motor neuron spikes to the joints: a first-order torque source, a series-elastic actuator, a pair of Hill-type muscles that can only pull, and a geared DC motor. We varied one property at a time (stiffness, maximum shortening velocity, curvature of the force-velocity curve, gear stiffness, rotor damping, free speed) and built five mixed bodies, with motors on some legs or joints and muscles on the rest. That gave 20 settings per connectome.

The reflex survived nearly all of them. We counted it as surviving if the median held ratio was below 0.9 and at least half of the eight joint and direction cases were reduced. MaleCNS kept the reflex in 20 of 20 settings, with ratios from 0.44 to 0.88. BANC kept it in 19 of 20, more weakly: 0.70 to 0.79 for most settings and 0.81 to 0.88 for the borderline ones, which we remeasured with ten replicates. The one failure put the motor on every femur-tibia joint and muscles everywhere else (0.92). With the motor on all joints the reflex survived (0.72), so the mismatch between neighbouring joints, not the motor itself, is the likely cause. An earlier sweep of the series-elastic actuator showed where the reflex ends. In MaleCNS it survived with low series damping ($b\le0.5$) and was lost at $b\ge2$. BANC was less tidy: two soft springs failed even at $b=0.1$, two stiff ones survived at $b=2$, and every setting failed at $b=10$.

Walking showed the opposite. Because the connectome does not walk, we drove the legs with a labelled stand-in, the hybrid controller of NeuroMechFly v2~\cite{wangchen2024}, commanded by three descending neurons: DNg100 for forward walking, the left-right difference of DNa02 for turning, and MDN for walking backward. We then sent the stand-in's joint targets as torque commands through each actuator. In this mode the leg motor neurons are not used; the connectome reaches the legs only through the descending command. Speed followed actuator bandwidth, not actuator kind. The first-order actuator walked at 11.1~mm/s in BANC and 14.0~mm/s in MaleCNS, as fast as the position-controlled reference. The Hill muscle pair walked at 4.6 and 5.0~mm/s, the DC motor at 5.4 and 6.4~mm/s, and a faster motor at 10.0 and 11.9~mm/s. The slow actuators did half the positive work per millimetre: about 44~µN against 89 to 91~µN for the first-order source. Visual steering, measured as the difference in heading between a drum turning left and one turning right, stayed within about 11 degrees for every actuator, the same size as the spread between seeds.

The energy ledger closed at every step. The actuators balanced to round-off: at most $5\times10^{-14}$ of the energy that passed through them at a physics step of 0.01~ms, and $2\times10^{-13}$ for the stiffest motor. The body balanced to within 0.19\% at 0.01~ms; the stiffest motor setting, a gear stiffness of 30,000, needed 0.0033~ms. At 0.1~ms the coupled system had not converged, and an apparent sideways drift of the DC motor was a numerical artefact.

\subsection*{Senses in the loop, and the whole fly}

\begin{figure}[!h]
\includegraphics[width=\textwidth,height=0.8\textheight,keepaspectratio]{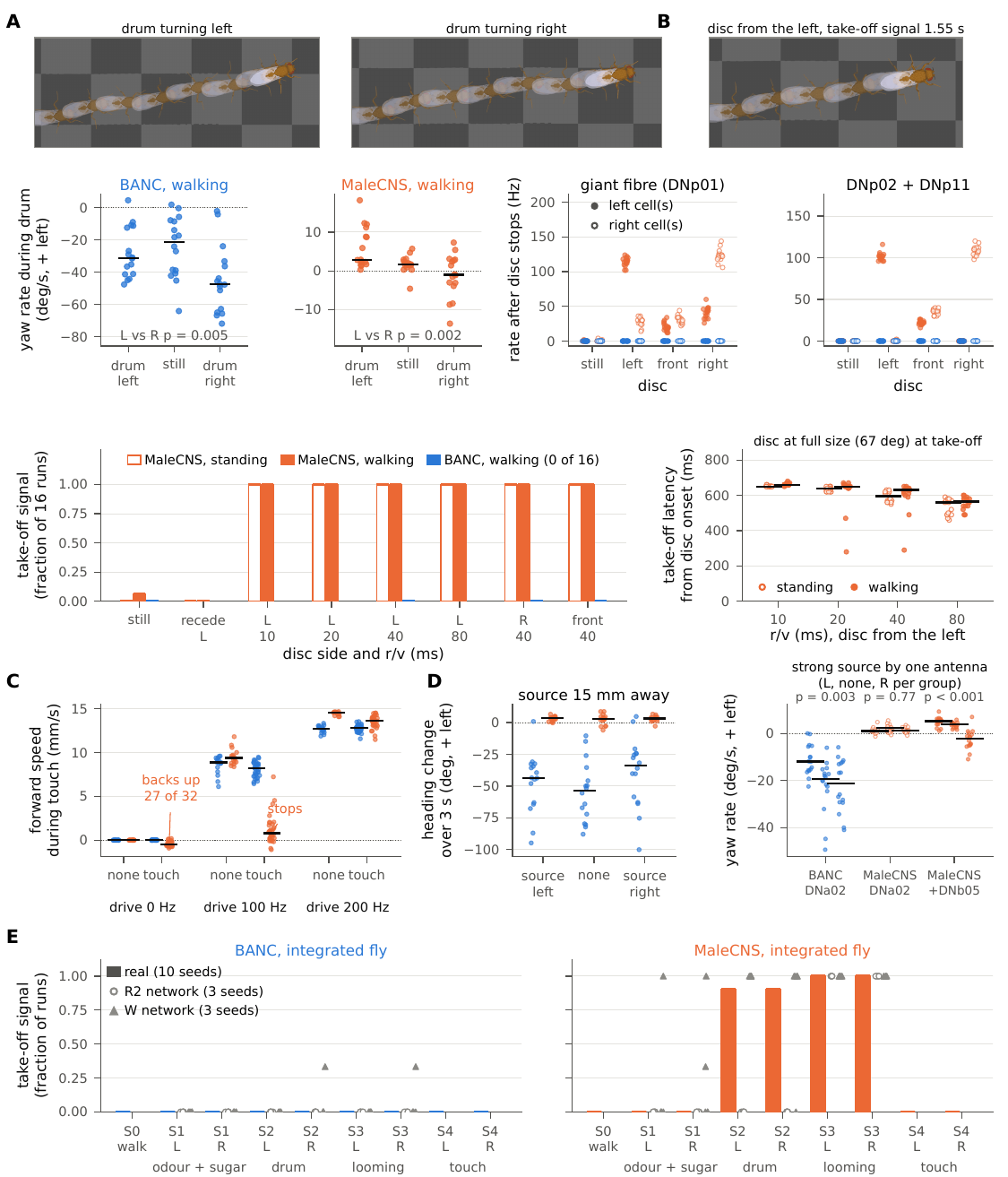}\par
\caption{\textbf{Senses steer the walking fly through the connectome.}
A: Drum steering. Top: the walking MaleCNS fly seen from above, 0.7 to 2.0~s (seed 0, 200~Hz drive), drawn every 200~ms; the view is rotated so the fly walks to the right, and a left turn bends up. The drum is not drawn. In this seed heading changed by 6.5 degrees with the drum turning left and by $-1.6$ degrees with it turning right, against 4.7 degrees with a still drum, too little to see in a still. Below: yaw rate per seed in the 900 to 1,500~ms drum window, walking at the 200~Hz drive, 16 seeds per condition; bars are medians; one-sided Mann-Whitney test, left against right drum. B: Looming. Top: the same view with a disc approaching from the left from 900~ms (not drawn); the detector signalled take-off at 1.55~s and the fly stopped, so the last positions overlap (forward speed 14.8~mm/s before the disc, 2.7~mm/s after it stops). Below: escape neuron rates per seed in the 500~ms after the disc stops (walking; DNp02 and DNp11 summed); take-off fraction with the two-spike detector for every condition, standing and walking, with approach speed as $r/v$; latency with the default single-spike detector. C: Antennal touch: forward speed during the probe, left and right antenna pooled ($n=32$) against no touch ($n=16$), at walk drives of 0, 100 and 200~Hz. Backing is a displacement below $-0.1$~mm. D: Odour. Heading change over 3~s with a source 15~mm away at $\pm90$ degrees; yaw rate with a strong source next to one antenna, with the labelled DNb05 arm (weight 0.75). E: The integrated fly: take-off fraction for each scenario and side. Bars, real network (10 seeds); points, each R2 and W network (5 each, 3 seeds), for scenarios S1 to S3 only. Drawn by a script written with Claude Code (see Use of AI tools).}
\label{fig4}
\end{figure}

Every sense reached the connectome, and some reached behaviour, with clear differences between the datasets (Fig~\ref{fig4}).

\textbf{Vision.} We rendered the fly's compound eyes~\cite{stavenga2003a,stavenga2003b}, ran them through the connectome-constrained optic lobe of flyvis~\cite{lappalainen2024}, and fed its columnar outputs to the connectome's own neurons of the same type and column. A drum turning left or right steered the walking MaleCNS fly with the drum in both directions ($p=0.002$, 16 seeds), by about 4 degrees per second. BANC steered only to the right. Its left eye and optic lobe are largely untyped in the dataset, so this is a data gap, not biology.

\textbf{Looming.} A disc expanding on one side of the walking MaleCNS fly drove the giant fibre at 114 to 123~Hz and DNp02 and DNp11 at 50 to 56~Hz on that side; in the standing fly the rates were somewhat higher. A labelled detector, which reads two giant fibre spikes within 10~ms, signalled take-off in 16 of 16 runs in each of 12 conditions. It never signalled for receding discs and at most once for still ones. With the default single-spike rule, take-off came only at full disc size, 565 to 660~ms after onset, with no tuning to approach speed. BANC never signalled take-off. With labelled one-way gap junctions, the jump motor neuron TTMn and the interneuron PSI followed every giant fibre spike 0.2~ms later in both connectomes, against about 0.3~ms in flies. The flight muscle motor neurons did not: they followed at most half the spikes, 4 to 5~ms later, in MaleCNS, and none in BANC.

\textbf{Touch and smell.} In MaleCNS, touch on one antenna made the fly back up when it had no walk drive (27 of 32 runs moved back by more than 0.1~mm). With a moderate walk drive it stopped (0.3 to 1.4 against 9.4~mm/s), and with the full drive it only slowed (13.5 to 13.9 against 14.5~mm/s). BANC did neither. An odour source 15~mm away did not steer either connectome: the two antennae differed in concentration by less than 1\%. A strong odour next to one antenna steered MaleCNS toward it only when DNb05 joined DNa02 in the turning command, a labelled arm. The published networks also needed the two edits before smell worked at all: without them about 7,900 BANC neurons sustained their own activity (Materials and methods).

\textbf{The whole fly.} We then turned every sense on at once, with the escape detector and the gap junctions, and ran five scenarios: walking, foraging toward odour and sugar, a drum, looming, and antennal touch. Combining senses produced one new interaction. MaleCNS signalled take-off to looming in 20 of 20 runs, and this survived all five R2 and all five W networks (30 of 30 runs each). The drum also triggered take-off in MaleCNS, 18 of 20 times; this survived W in 9 of 10 network and side pairs but not R2. Some W networks signalled take-off with no threat at all. The foraging fly did not steer to odour, touch did not slow it at the full walk drive, and sugar under its feet never reached MN9.

\subsection*{Flight from motor neuron spikes}

\begin{figure}[!h]
\includegraphics[width=\textwidth,height=0.8\textheight,keepaspectratio]{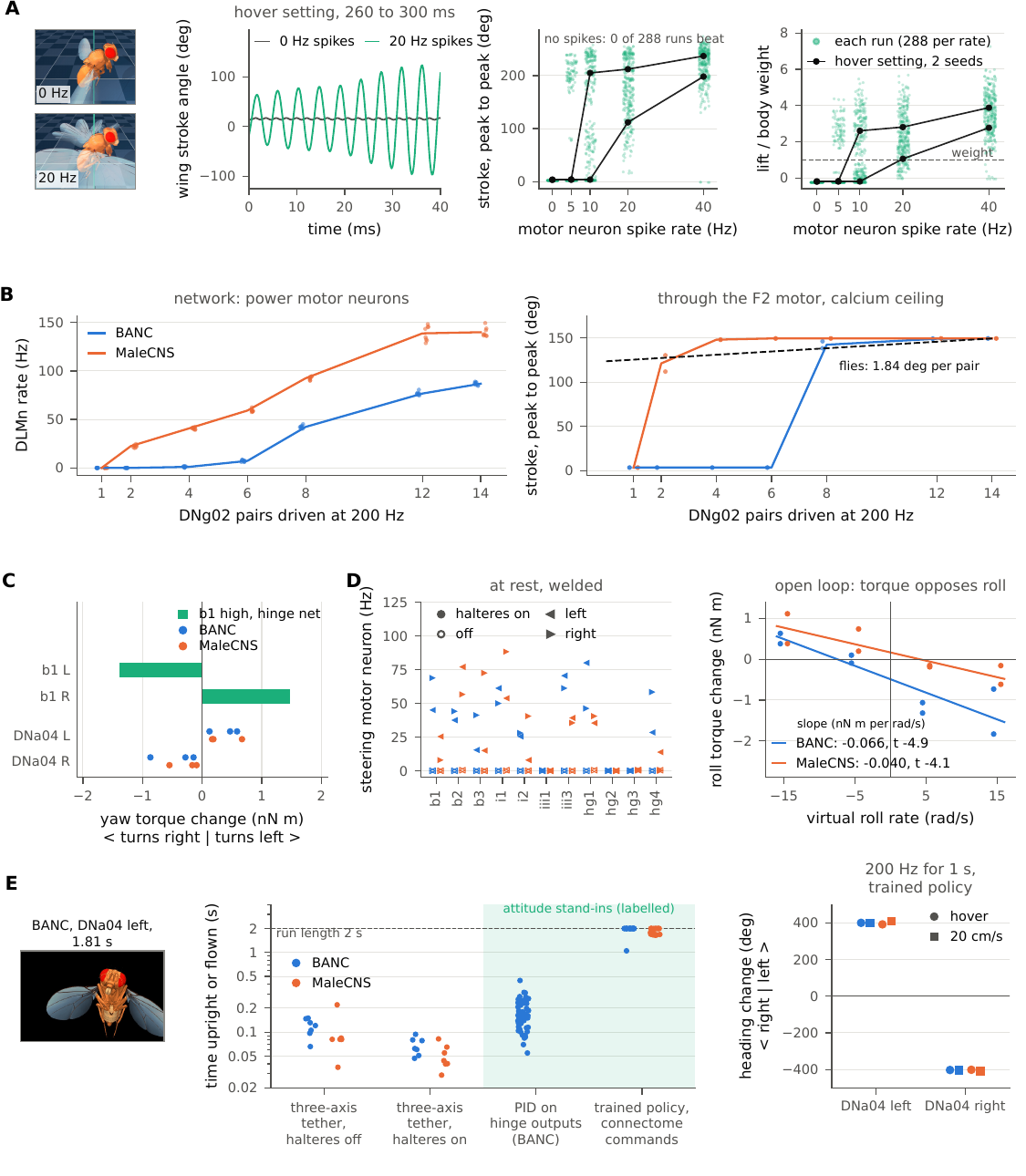}\par
\caption{\textbf{Flight from motor neuron spikes.}
A: The asynchronous power muscle beats the wings only when its motor neurons spike. Far left: stroboscopic stills of the tethered fly over one wingbeat, 195.0 to 199.4~ms (12 frames 0.4~ms apart, combined by pixel maximum), at 0 and 20~Hz drive (hover setting, seed 0); at 0~Hz the wings stay still, at 20~Hz the wing sweeps a fan. Stroke trace at the hover setting with 0 and 20~Hz drive (seed 0, mean of both wings); stroke amplitude and lift against spike rate for all 1,440 runs of the muscle sweep (0.3~s, Poisson drive, last 100~ms), with the two seeds of the hover setting as lines. B: DNg02 recruitment of the power motor neurons at 200~Hz drive, both connectomes (network, 8 seeds), and the stroke it produces through the muscle with the calcium ceiling (2 seeds); dashed line, flies' 1.84 degrees per pair. C: Steering through the wing hinge network: yaw torque from b1 asymmetry (b1 at 1 on one side, 0 on the other) and from DNa04 at three stimulation rates; positive turns left. D: Haltere signals: steering motor neuron rates with halteres on and off (mean of 2 seeds), and roll torque against imposed roll rate with least-squares fits. E: Far left: the BANC fly under the trained policy at 20~cm/s, 1.81~s, with DNa04 on the left driven at 200~Hz from 1 to 2~s (seed 0). Time upright on the tether (halteres off and on), with the labelled hinge controllers and with the trained policy; points are condition means. Tether and controllers: time to 45 degrees; policy: time flown. Heading change under DNa04 stimulation (200~Hz, 1 to 2~s), mean and SD of 3 seeds. Drawn by a script written with Claude Code (see Use of AI tools).}
\label{fig5}
\end{figure}

Flight muscles are of two kinds. The power muscles are asynchronous: stretch-activated, they beat the wings at about 200~Hz while their motor neurons fire far more slowly~\cite{pons2023}. The steering muscles are synchronous. We built both into the flight body of flybody~\cite{vaxenburg2025} (Fig~\ref{fig5}).

\textbf{Power.} A port-Hamiltonian stretch-activated muscle pair, driven only by motor neuron spikes, beat the wings of the tethered fly by itself. With no spikes nothing moved: none of 288 runs oscillated. Spike rate acted as a throttle. In 0.3~s runs with a 20~Hz Poisson drive, lift reached one to three times the fly's weight, and hover cost 133 to 166~W per kg of muscle, against 171 for the trained reference. These short runs were partly transient: the two seeds of the hover setting differed by a factor of two. A constant 12~Hz drive then held lift at the fly's weight (0.99 to 1.01 times). The energy of the motor closed within 0.34\% (median). A labelled ceiling on muscle calcium, used from then on, capped the stroke at 149 degrees and lift at 1.39 times weight; with it, 12~Hz gave 0.71 times weight.

\textbf{Throttle.} Descending neuron DNg02, which flies use to set wing amplitude~\cite{namiki2022}, recruited the power motor neurons in both connectomes past a threshold: about six pairs in BANC and two in MaleCNS at 200~Hz. Recruitment was steep. Through the muscle model each added pair raised the stroke by 23 to 49 degrees, against 1.84 in flies, so the throttle was nearly on-off. With halteres in the loop it became graded: whenever the wings beat, BANC's DNg02 drive set lift between 0.3 and 1.4 times weight. Under W the throttle halved.

\textbf{Steering.} We drove the steering muscles through the wing hinge network of Melis et al.~\cite{melis2024}, ported exactly from TensorFlow. Raising muscle b1 on one side yawed the fly away from it, symmetrically. Single-muscle effects matched the published signs for 8 of 11 muscles on stroke and 7 of 11 on deviation. Stimulating descending neuron DNa04 turned the fly toward its own side in both connectomes.

\textbf{Halteres.} We added halteres as gyroscopic sensors: a beating mass, Coriolis shear, and afferents that spike at fixed phases. They drove the steering motor neurons at rest in both connectomes, and in open loop they opposed roll ($t=-4.9$ in BANC, $-4.1$ in MaleCNS). In closed loop they did not correct perturbations. The body model damped rotation faster than flies do. In free flight it halved body rotation within 0.55 wingbeats; on a rotating tether, yaw took 1.84 wingbeats and roll 0.63, against about two in flies~\cite{hedrick2009}.

\textbf{Stability.} With the stretch-activated muscle, the hinge network and halteres, the free fly tipped past 45 degrees within 28 to 245~ms. A labelled controller on the hinge outputs kept it up for at most 0.54~s. Only flybody's trained flight policy, a labelled stand-in, flew. The connectome then commanded it: DNa04 turned the fly by about 400 degrees per second toward its own side in both connectomes, and doubling the DNg02 drive made BANC climb 3~cm. No descending neuron in the command responded to body perturbations, and a turning drum did not steer flight.

\subsection*{Two animals, specific differences}

\begin{figure}[!h]
\includegraphics[width=\textwidth,height=0.8\textheight,keepaspectratio]{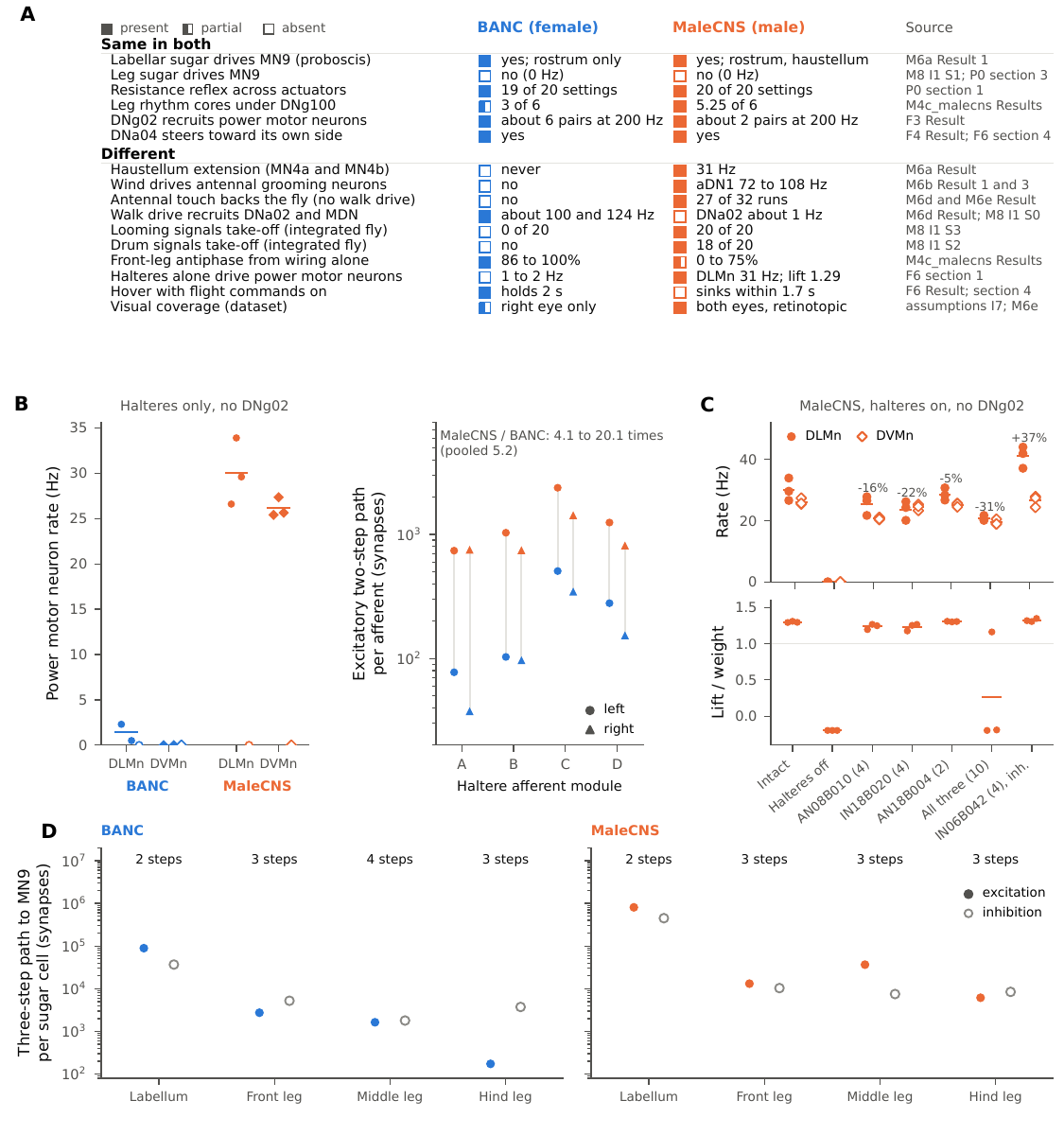}\par
\caption{\textbf{Two animals, specific differences.}
A: Results compared between BANC and MaleCNS: the same in both (6 rows) and different (10 rows). Filled square, present; half, partial or weaker; open, absent; report section at right. B: Left: power motor neuron rates with halteres on and no DNg02 input, welded body, 1~s; points are seeds (BANC $n=2$, MaleCNS $n=3$), bars means, open markers halteres off. Right: excitatory two-step synapses per haltere afferent for modules A to D, left (circles) and right (triangles), log scale. C: Silencing relays in MaleCNS: DLMn (filled) and DVMn (open) rates and lift per seed ($n=3$) for each silenced type, with cell counts; percentages are the DLMn change from intact. D: Three-step excitation (filled) and inhibition (open) per sugar cell, log scale, BANC and MaleCNS side by side; above each organ, the fewest excitatory steps to MN9 (connections of at least 5 synapses). Drawn by a script written with Claude Code (see Use of AI tools).}
\label{fig6}
\end{figure}

The two connectomes agreed on feeding, the reflex, DNg02 recruitment and DNa04 steering, and in kind on the leg rhythm. They disagreed on much else (Fig~\ref{fig6}). MaleCNS backed up to antennal touch without a walk drive, signalled take-off to looming, extended the haustellum and recruited its antennal grooming neurons in wind; BANC did none of these. BANC's walk drive recruited the turning neuron DNa02 at about 100~Hz and the backward neuron MDN at 124~Hz, which made the stand-in drift; MaleCNS's did not. Some differences come from the data: BANC's left visual system is incomplete. Others trace to wiring.

The clearest case was a haltere drive of the flight power muscles. In MaleCNS, halteres alone, with no DNg02 input, drove the power motor neurons at about 30~Hz (DLMn) and 26~Hz (DVMn) and lifted the tethered fly at 1.3 times its weight. BANC's halteres did not: its DLMn fired at 0.5 to 3.3~Hz, its DVMn not at all, and the wings stopped. Excitatory two-step paths from haltere afferents to the power motor neurons were 4 to 20 times stronger per afferent in MaleCNS, depending on the afferent module (5.2 times pooled). Silencing single relays confirmed the paths and their signs. Removing the outputs of AN08B010 lowered the DLMn rate by 16\%. Removing three excitatory relays together lowered it by 31\% and stopped the wings in two of three seeds. Removing the inhibitory relay IN06B042 raised it by 37\%. No single relay carries the drive; it is spread over the nerve cord.

The second case was a negative one. Sugar under the feet never drove MN9, in either connectome, although flies extend the proboscis to tarsal sugar. Labellar sugar neurons reach MN9 in two steps. Leg sugar neurons need three or four, and their three-step excitation per cell is 20 to 500 times weaker; in BANC, inhibition outweighs excitation for every leg. Every leg sugar neuron has outputs, so the data are complete. A network with synapse counts as weights cannot carry this signal.

\subsection*{A published closed loop, rebuilt}

\begin{figure}[!h]
\includegraphics[width=\textwidth,height=0.8\textheight,keepaspectratio]{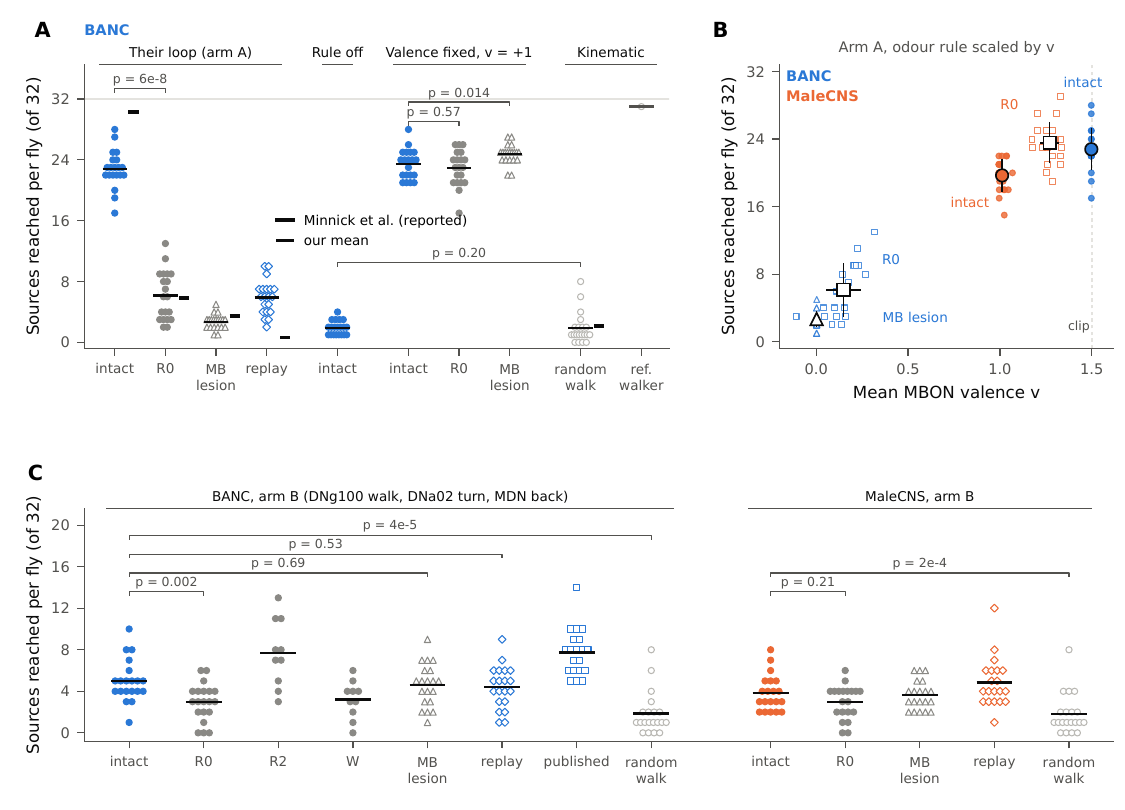}\par
\caption{\textbf{A published closed loop, rebuilt: the wiring enters through one number.}
A: Food sources reached per fly (of 32; 9 arenas of 20~s) when BANC drives the rebuilt loop of Minnick et al.: intact network, R0, mushroom-body lesion (Kenyon cells and MBONs removed) and open-loop replay from a start rotated by 30 degrees; the loop with its odour rule off; the loop with MBON valence fixed at $v=+1$ (intact, R0, lesion); kinematic controls at the loop's median speed (random walk; reference walker heading to the nearest source, $n=1$). Thin bars, our means; thick black bars, values reported by Minnick et al. B: Sources reached against mean MBON valence (approach minus avoidance MBON spikes over a normaliser, clipped at $\pm1.5$), per fly (small symbols) and per condition (large, $\pm$SD). BANC uses the paper's fixed normaliser; MaleCNS a normaliser calibrated on each network. C: Our descending-neuron stand-in with no hand-written rule: BANC default network, R0, R2, W, lesion, replay and published network; MaleCNS intact, R0, lesion and replay; random walk. $n=20$ flies per condition (10 for R2 and W); two-sided Mann-Whitney tests. Drawn by a script written with Claude Code (see Use of AI tools).}
\label{fig7}
\end{figure}

Minnick et al. ran BANC's whole nervous system in NeuroMechFly and reported odour navigation that shuffled networks could not reproduce~\cite{minnick2026}. Their result bears directly on our question, so we rebuilt their loop from the paper and ran it on our pipeline (Fig~\ref{fig7}). In their loop, leg motor pools set the amplitude of each leg's oscillator, and a hand-written rule, scaled by a valence computed from mushroom body output neurons (MBONs), slows the legs on the side with more odour.

Our rebuild did not fully reproduce their score. The intact BANC fly reached 22.8 of 32 food sources against their 30.25, which misses the fidelity criterion we set in advance. The pattern did reproduce. Shuffled networks (R0) reached 6.2 sources, as in their paper, a lesion of Kenyon cells and MBONs 2.7, and a random walk 1.9 ($p<10^{-7}$ for each against intact, 20 flies each).

The navigation came from the rule. With the odour rule switched off, the intact fly reached 1.9 sources, the same as a random walk ($p=0.20$), although its network still produced the full valence. The network entered the loop through that one number, and success followed its size: 1.50 in the intact network, 0.15 under R0 and 0 after the lesion (Fig~\ref{fig7}B). In MaleCNS, whose valence we normalised per network, R0 kept a large valence (1.27) and beat the intact network (23.6 against 19.7 sources, $p=10^{-5}$). We then fixed the valence at $+1$ in BANC. All three networks navigated equally: 23.5 sources intact, 22.9 under R0 ($p=0.57$) and 24.8 after the lesion, slightly more ($p=0.014$). In this loop, the dependence on wiring is the sign and size of one scalar, not a navigation circuit.

Our own stand-in, which steers only through descending neurons and has no hand-written rule, barely navigated: 5.0 sources in BANC and 3.9 in MaleCNS. BANC's score beat R0 and the random walk, but its open-loop replay did as well (4.4, $p=0.53$); in MaleCNS the replay reached 4.9 ($p=0.12$). The advantage came from the shape of the path, not from sensing: our stand-in does not navigate with its senses.

\section*{Discussion}

We put two whole-CNS connectomes into simulated bodies and asked what the wiring does there. The answer splits cleanly. The wiring gives local functions: a reflex at a joint, a rhythm in a leg, a feeding command, an escape signal, a throttle and a steering command for flight. It does not give the coordination that turns these into behaviour: six legs that walk, a body that stays upright in the air, a correction after a push. For those we needed engineered stand-ins. The connectome could command them, but it could not replace them.

\subsection*{The reflex lives in the wiring, the gait in the actuator}

The cyborg experiments separate two kinds of property. The resistance reflex survived 39 of 40 actuator settings across the two connectomes, from an ideal torque source to a pull-only muscle pair and a geared motor. With port-Hamiltonian actuators, the same reflex fell outside every null network that kept cell types but moved partners or synapse counts. So the reflex is set by who connects to whom, and it tolerates large changes in how the force is made. Walking speed and energy cost behaved the other way. Under the walking stand-in they followed actuator bandwidth, from about 5 to 14~mm/s, while the connectome's visual steering did not change. That speed result belongs to the stand-in and the actuator; the connectome only issued the command.

This split suggests a practical rule for neuromechanical models. Wiring-level functions can be tested with simple actuators, because they barely depend on them. Gait, speed and cost cannot; they need actuators whose bandwidth and energetics are right. The port-Hamiltonian ledger made this comparison possible. It closed to round-off in every actuator and within 0.19\% in the body, so energy differences between actuators are real, not numerical.

\subsection*{What is missing for coordination}

Why does the wiring give a rhythm in each leg but not a gait? The leg circuits oscillated only in a rate model with a smooth command, as in Pugliese et al.~\cite{pugliese2026}; spiking versions did not. Even then, six legs did not coordinate. Three explanations fit. Synapse counts are not weights: connection strengths vary within and across animals~\cite{berg2026}, and the null networks show that our results depend on counts (W) as well as on partners. Neuromodulation, plasticity and most gap junctions are absent from the connectome and from our model. And the body model's passive properties may differ from the fly's: in free flight our model halved body rotation within about half a wingbeat, against about two wingbeats in flies~\cite{hedrick2009}, which changes what a controller must do.

Flight stability failed for related reasons. The stretch-activated motor beat the wings and the hinge network steered them, but no descending signal responded to a perturbation. Halteres drove the steering motor neurons and opposed roll in open loop, yet closing the loop did not correct pulses. A trained policy flew, and the connectome commanded it well: DNa04 turned the fly toward its own side in both animals, as the probe predicted. We read this as evidence that the connectome carries the command layer of flight. The fast stabilising loop either lives in properties the wiring does not record, such as gap junctions, or needs a body model closer to the fly.

\subsection*{Two animals}

Comparing BANC and MaleCNS was the most useful check we had. It is not a test of sex differences: one animal of each sex cannot separate sex from individual variation or reconstruction differences, which are large even within one circuit~\cite{yad2026}. It is a test of robustness. Results that held in both connectomes, such as feeding, the reflex, DNg02 recruitment and DNa04 steering, are less likely to depend on one reconstruction. Results that held in one, we traced where we could. The haltere drive of the power muscles in MaleCNS came from excitatory two-step paths 4 to 20 times stronger per afferent than in BANC, spread over several nerve cord relays whose silencing changed the drive with the predicted signs. Haltere afferents reach the steering motor neurons in a female nerve cord~\cite{dhawan2026}; the path to the power muscles is, to our knowledge, new. Whether flies use it, and whether it differs between sexes, is an experimental question.

\subsection*{Stand-ins can make wiring look necessary}

Embodying a fly connectome is no longer new. Amandeep ran both connectomes with nested null networks in a walking body~\cite{amandeep2026}, and Aldunate Rodr\'iguez listed behaviours that do and do not emerge from MaleCNS~\cite{aldunate2026}. Minnick et al. ran BANC's whole nervous system in NeuroMechFly with the same neuron model and weight, and reported odour navigation that rewired networks could not reproduce~\cite{minnick2026}. In a faithful rebuild of their loop on our pipeline, the pattern held but the score did not: 71\% of sources against their 95\%. The navigation came from their hand-written odour rule, and the network entered through one number, the MBON valence. With that number fixed, shuffled and lesioned networks navigated as well as the intact one.

We do not read this as a flaw peculiar to one study. Any rule placed between the connectome and the body can turn a single readout into behaviour, and a null network then tests the readout, not a circuit. Our own walking stand-in has the same structure, through three descending neurons, which is why we report every result with the stand-in named and, where possible, against its open-loop replay. Under that control our stand-in did not navigate to odour at all. The general lesson is that a null-network test of an embodied model is only as specific as the interface it runs through.

Against this background our work adds four things: actuators that can be swapped under a fixed connectome with an exact energy audit; flight driven by motor neuron spikes through an asynchronous muscle; null networks applied to body-level reflexes, the flight throttle and energy, including synapse-count shuffles and shuffles that keep the leg segment; and a direct test of a published embodied result.

The Digital Sphinx critique asked whether the wiring matters, since a worm connectome with a learned decoder can steer a fly body~\cite{brunton2026}. Our design answers in part. We learned no decoder: the interface from motor neurons to actuators is fixed and stays fixed across swaps. Null networks then show which results need the real wiring. Feeding, the reflex, the rhythm, the throttle and MaleCNS's looming take-off did. The drum's effect on walking did not separate from R2 or W, and we do not claim it as wiring-specific.

\subsection*{Limitations}

Synapse counts stand in for weights, with one weight per dataset. Gap junctions are absent except for labelled giant fibre and haltere junctions, and neuromodulation and plasticity are absent. With five null networks per level, a result can fall outside the null range by chance one time in three, so single-level results are indications; the strongest claims rest on agreement across levels and connectomes. Vision is replayed from a still fly rather than rendered live, smell uses a static plume, and leg proprioception is off under the walking stand-in. Several muscle parameters come from other insects, and the map from motor neurons to muscles comes from the literature, not the connectome. Walking and flight use different body models, so the fly cannot take off. MaleCNS, from a male, drives a female body. Our rebuild of the published navigation loop fills gaps in its description with our own choices. Each of these choices is labelled in the assumption register (S1 Table) and can be switched off.

\subsection*{Conclusion}

A whole-CNS connectome, run as spiking neurons in a body, carries reflexes, rhythms and commands, and these survive when engineered actuators replace muscles. Coordination does not emerge from the wiring with these models. This gives a clear division of labour for building a synthetic fly: take the commands and reflexes from the connectome, and engineer, measure and label what it does not give.

\section*{Materials and methods}

\subsection*{Connectomes}
BANC v888~\cite{bates2026} was pinned to Harvard Dataverse version 3.0~\cite{bates2026data}, with synapse release v3 (synapses of size 10 or more) by default and v2 kept for comparison. A neuron was a proofread or roughly proofread cell, excluding glia, trachea and fragments labelled not a neuron: 155,858 cells against 155,916 in the paper. MaleCNS v1.0~\cite{berg2026} was read from its public flat files; a neuron was a body with a superclass label (166,700). Removing autapses left 13,366,670 directed connections and 41,845,636 synapses in BANC, and 25,582,837 connections and 124,177,143 synapses in MaleCNS. Every file was downloaded over HTTPS with its checksum recorded. We grouped the curators' labels into sensory neurons by organ and modality, motor neurons by nerve and body part, descending, ascending and intrinsic neurons, without changing any label [I1]. The networks hold 16,151 sensory and 805 motor neurons in BANC, and 17,937 and 815 in MaleCNS.

\textbf{Sign rule.} A neuron is inhibitory if more than half of its presynaptic sites are predicted GABA or glutamate, and excitatory otherwise, including neurons with no prediction. This makes 71.9\% of BANC nodes excitatory. The connection from $j$ to $i$ has weight $W_{ij}=\pm n_{ij}w_{\mathrm{syn}}$, with $n_{ij}$ the synapse count and the sign of neuron $j$.

\textbf{Labelled network edits.} From the senses onward we used two edits by default, both switchable. The first is a sign fix: BANC's APL neuron has no transmitter prediction but a verified GABA label, so the rule above made it excitatory; we made it inhibitory [N3]. The second removes a pathway: the chemical outputs of excitatory antennal-lobe local neurons onto local and projection neurons, 36,261 connections and 354,359 synapses in BANC [N13]. In flies these neurons excite projection neurons largely through gap junctions, which the model would otherwise turn into one-way chemical kicks. Without the two edits, odour input started self-sustained activity in about 7,900 BANC neurons. For escape and flight in BANC we also made the two PSI neurons excitatory, since they are predicted glutamatergic but drive the flight muscle motor neurons in flies [N16]. Recent work supports the APL fix~\cite{minnick2026}.

\subsection*{Neuron model}
Each neuron follows the leaky integrate-and-fire model of Shiu et al.~\cite{shiu2024}:
\begin{eqnarray}
\label{eq:lif}
\tau_m \dot v = v_0 - v + g, \qquad \tau_s \dot g = -g,
\end{eqnarray}
with $v_0=-52$~mV, threshold $-45$~mV, reset to $-52$~mV with $g=0$, membrane time constant $\tau_m=20$~ms, synaptic time constant $\tau_s=5$~ms, a refractory period of 2.2~ms and a synaptic delay of 1.8~ms. A spike of neuron $j$ adds $W_{ij}$ to $g_i$ after the delay. We integrate Eq~(\ref{eq:lif}) exactly over each step of $\Delta t=0.1$~ms, then apply threshold, spike delivery and reset in the order Brian2 uses. Synaptic input that arrives during the refractory period is lost, as in the published model. External drive is a Poisson process: in each step a stimulated neuron receives an event with probability $r\Delta t$, and each event adds $250\,w_{\mathrm{syn}}$ to $v$; as in the published model, stimulated neurons have no refractory period [N2]. The weight per synapse is 0.275~mV in BANC and 0.18~mV in MaleCNS, the largest values that kept the sugar response selective [N9]. Our solver is written in NumPy and SciPy. It matched Brian2 2.10.1 and flycns 0.7.0 spike for spike, and gave identical spikes on Windows and Linux.

\textbf{Leg rhythm.} For the leg rhythm we used the rate model of Pugliese et al.~\cite{pugliese2026},
\begin{eqnarray}
\label{eq:rate}
\tau_r \dot r = -r + \max\!\left(r_{\max}\tanh\!\left(\frac{a}{r_{\max}}(I + Wr - \theta)\right), 0\right),
\end{eqnarray}
with per-neuron parameters drawn from normal distributions truncated at zero ($\tau_r$ $20\pm2$~ms, $a$ $1\pm0.1$, $\theta$ $7.5\pm0.6$, $r_{\max}$ $200\pm10$~Hz). Here $W$ is the signed synapse count times $b=0.03$; in MaleCNS, whose neurons receive two to three times more synapses, $b$ was scaled by 0.5 or 0.65. Excitability was scaled by each neuron's surface area $s$, as $a\to a/s$ and $\theta\to\theta s$ [RH2, RH15]. The descending command was a smooth level rather than spikes [RH8]. A leg core counted as oscillating if its motor neurons had an oscillation score above 0.5 at 5 to 25~Hz. The level of the command and the treatment of BANC's fragment loop differ between the rhythm study and the null-network test; each result names its setting.

\textbf{Gap junctions.} The connectome records only chemical synapses. In escape experiments, and in the integrated walking fly, we added labelled one-way junctions: each giant fibre spike raised the membrane potential of its own-side TTMn and the PSI neurons it contacts by 10.5~mV, 0.1~ms later [E2]. In flight, the module C haltere afferents excited the b1 steering motor neuron on their side by 10~mV in the same way [FL12]. A faster PSI to DLMn link was tested only in the escape diagnosis [E3]; no flight run used it, because no flight run starts from a giant fibre spike.

\subsection*{Ports and co-simulation}
A port has a neural side and a physical side. A sensor port maps body state to firing rates of named sensory neurons, which are converted to Poisson events every 0.1~ms; sensor values are read every 1~ms. An actuator port maps spikes of named motor neurons to a torque $\tau$ at a joint with angle $q$ and angular velocity $\omega=\dot q$, so its delivered power is $P=\tau\omega$. Models are chosen by name in the run's configuration file, so swapping an actuator changes no code.

\textbf{Port-Hamiltonian form.} Every actuator is written as an input-state-output port-Hamiltonian system~\cite{maschke1992,vanderschaft2017,vanderschaft2014},
\begin{eqnarray}
\dot x = \left[J(x)-R(x)\right]\frac{\partial H}{\partial x}(x) + g(x)\,u, \qquad y = g(x)^{\top}\frac{\partial H}{\partial x}(x),
\end{eqnarray}
with state $x$, Hamiltonian $H(x)\ge 0$ (the stored energy), a skew-symmetric interconnection matrix $J(x)=-J(x)^{\top}$, a symmetric positive semi-definite dissipation matrix $R(x)=R(x)^{\top}\succeq 0$, an input matrix $g(x)$, and an input $u$ and output $y$ whose product is power. Along solutions,
\begin{eqnarray}
\dot H = -\frac{\partial H}{\partial x}^{\top} R(x)\,\frac{\partial H}{\partial x} + y^{\top}u \;\le\; y^{\top}u ,
\end{eqnarray}
so an actuator can store and dissipate energy but cannot create it: whatever it delivers enters through its ports~\cite{ortega2001,ortega2002}. Each leg actuator has two ports. The source port carries the command set by the motor neurons; its power is the energy drawn from the actuator's supply, which plays the role of metabolism. The joint port carries the pair $(\tau,\omega)$. A power-conserving interconnection of port-Hamiltonian systems is again port-Hamiltonian, with the sum of the Hamiltonians as stored energy~\cite{vanderschaft2014,duindam2009}. This is what makes actuators interchangeable: any model with the same two ports can be connected to the same motor neurons and the same joint. We have used this formulation to model and control robot manipulators~\cite{ma2026}, and have reviewed its use outside engineering, including biology~\cite{munozarias2026}. The body is not port-Hamiltonian: MuJoCo integrates its rigid-body dynamics, and we audit its energy from outside.

Each actuator keeps a ledger of energy taken from its source $W_{\mathrm{in}}$, returned to it $W_{\mathrm{out}}$, dissipated $D$ and delivered to the body $W_{\mathrm{del}}$. With stored energy $H$, the residual $H-(W_{\mathrm{in}}-W_{\mathrm{out}}-D-W_{\mathrm{del}})$ is checked at every recorded step. We report it relative to the energy that passed through the actuator. The body balance compares the change in MuJoCo's mechanical energy with motor, external and contact work minus joint damping. The actuator books $\tau\omega$ with $\omega$ held over a physics step, while the body integrates with $\omega$ changing within it, so the two sides of the joint port agree only to first order in the step; this is one reason the body balance closes less tightly than the actuator balance.

A scheduler steps the components at multiples of a 0.1~ms base tick. In each tick, sensors are read (every tenth tick), the network takes one step, actuators read this step's spikes, and actuator and physics take $n$ substeps with spikes held. Torques act one physics step after the spikes that caused them. From the actuator study on, the physics step was 0.01~ms ($n=10$); at 0.1~ms the coupled system had not converged [P17]. Walking under the position-controlled stand-in ran at 0.1~ms.

\subsection*{Sensors}
\textbf{Leg proprioception and load.} Femur-tibia chordotonal neurons encode joint angle (claw), velocity in a preferred direction (hook) and speed (club); hair plates encode coxa protraction; campaniform sensilla encode contact force, saturating at one third of body weight. Each saturates at 200~Hz; this rate is a guess, and at 100~Hz no reflex appeared [P5, P7, P13].
\textbf{Taste.} Gustatory neurons of the labellum and tarsi fire at up to 100~Hz when their patch touches a food patch [P21].
\textbf{Smell.} Olfactory receptor neurons respond to concentration at each antenna through a saturating function weighted by glomerular response profiles from DoOR [P22, P23].
\textbf{Johnston's organ.} Each antenna is a port-Hamiltonian rotational oscillator with angle $q_a$, momentum $p_a$ and inertia $I_a$, $H=p_a^2/2I_a+kq_a^2/2$, $J=\left(\begin{smallmatrix}0&1\\ -1&0\end{smallmatrix}\right)$, $R=\mathrm{diag}(0,c)$ and the air torque as input, so $\dot H=\tau_{\mathrm{air}}\dot q_a-c\,\dot q_a^2$; it has a resonance of 200~Hz and quality factor 1.3, driven by air speed and sound; five neuron groups read its band-passed or tonic motion [P25, P26].
\textbf{Head touch.} Bristle and touch neurons of the antennae, eyes and head are recruited by contact with leg or probe tips and adapt with a 277~ms time constant [P27].
\textbf{Vision.} A compound eye samples space through ommatidia whose visual axes set its resolution~\cite{munozarias2022}. FlyGym's compound eyes~\cite{wangchen2024} were rendered and passed to the flyvis optic lobe~\cite{lappalainen2024}. The activity of each flyvis cell type and column became the firing rate of the connectome neurons of the same type and column, whose own connectome inputs were removed [V1, V3]. Stimuli were rendered with the fly still and replayed in time.
The exact tuning curves of every sensor are given in S1 Table.

\subsection*{Actuators}
All leg actuators share the motor neuron map: motor neurons are assigned to joints and directions from BANC's functional labels and measured geometry [P1]. Forces and torques are in µN and µN\,mm.
\textbf{First order} [P3, P4]. Each spike raises activation $a$ by $0.4(1-a)$, and $a$ decays with $\tau_a=12$~ms~\cite{azevedo2020}; joint torque $\tau_m$ is the force-weighted sum of activations, up to 30~µN\,mm per joint and direction. Between spikes it is integrated exactly.
\textbf{Series elastic} [P16]. A spring of stiffness $k$ and a damper $b$ in series between the first-order torque $\tau_m$ and the joint, with optional parallel spring $k_p$ and damper $d_p$. The state is $x=(d,q)$, with $d$ the twist of the series spring and $q$ the joint excursion from rest. In the form above, $J=0$, $R=\mathrm{diag}(1/b,\,0)$, $g=\left(\begin{smallmatrix}1/b & -1\\ 0 & 1\end{smallmatrix}\right)$ and $u=(\tau_m,\omega)$; the joint receives $\tau=kd-k_pq-d_p\omega$, the parallel damper acting directly on the joint port. With $\omega_a=(\tau_m-kd)/b$ the damper velocity, the balance reads
\begin{eqnarray}
H=\tfrac12 k d^2+\tfrac12 k_p q^2, \qquad \dot H = \tau_m \omega_a - b\, \omega_a^2 - d_p \omega^2 - \tau \omega ,
\end{eqnarray}
source power minus dissipation minus power delivered to the joint, with $k=300$~µN\,mm/rad and $b=0.5$ by default.
\textbf{Hill-type muscle pair} [A3-1 to A3-5]. Two muscles per joint that can only pull. Each has a contractile element and a linear series elastic element of stiffness $k$ and stretch $\ell_s$, the only storage, with $H=\tfrac12 k\ell_s^2$ and $\dot\ell_s=u_\ell+v$, where $v$ is the shortening velocity of the contractile element and $u_\ell=-s\,\omega$ the lengthening rate of the whole muscle ($s=\pm1$, the direction it pulls). The contractile element is an algebraic relation between force and velocity,
\begin{eqnarray}
F = F_0 f_V(v) - c v = k \ell_s, \qquad f_V = \frac{1 - v/v_{\max}}{1 + v/(\kappa v_{\max})} \ \text{(shortening)},
\end{eqnarray}
where $F_0=f_L(l)\,F_a$ is the isometric force, $F_a$ the first-order activation force and $f_L$ a Gaussian force-length curve. When lengthening, $f_V$ rises to at most 1.5. We used $v_{\max}=25$~rad/s and curvature $\kappa=0.15$~\cite{eldred2010}. The ratio of moment arm to fibre length (0.28~mm over 1.41~mm) follows stick insect data~\cite{guschlbauer2007}; their force-length curve has a plateau with near-linear sides, which we approximate with a Gaussian of width 0.3 in normalised length; the series stiffness, 300~µN\,mm/rad, is a guess [A3-4, A3-5]. Its power splits as $Fv = F_0v-\left[F_0(1-f_V)v+cv^2\right]$: a modulated source $F_0v$, the only way active power enters, and a nonlinear resistor whose dissipation $F_0(1-f_V)v+cv^2$ is never negative, since $f_V\le1$ when shortening and $f_V\ge1$ when lengthening. Hence $\dot H = F_0v - F_0(1-f_V)v - cv^2 - sF\omega$, and the muscle cannot make energy.
\textbf{Geared DC motor} [A4-1]. The state is $x=(\lambda,p,\theta)$: winding flux, rotor momentum and gearbox twist, with current $i=\lambda/L$, rotor speed $\omega_r=p/J_r$ and joint torque $\tau=k_g\theta$. With motor constant $K$, winding resistance $R_w$ and rotor damping $b_r$ (all reflected to the joint side),
\begin{eqnarray}
H=\frac{\lambda^2}{2L}+\frac{p^2}{2J_r}+\frac{k_g\theta^2}{2}, \quad
J=\left(\begin{smallmatrix}0&-K&0\\ K&0&-1\\ 0&1&0\end{smallmatrix}\right), \quad
R=\mathrm{diag}(R_w,\,b_r,\,0), \quad
g=\left(\begin{smallmatrix}1&0\\ 0&0\\ 0&-1\end{smallmatrix}\right),
\end{eqnarray}
with input $u=(V,\omega)$, $V$ the winding voltage, so that $\dot H = Vi - R_w i^2 - b_r\omega_r^2 - \tau \omega$. The voltage is $V=R_w\tau_m/K$, clipped at the full command,
with constants set from free speed (60~rad/s), electrical (0.1~ms) and mechanical (2~ms) time constants, and stall torque equal to the command. By default the gear stiffness is $k_g=3000$~µN\,mm/rad and the rotor damping $b_r=0$.
\textbf{Mixed bodies} [MX-1] give each joint its own model and ledger.
The storing actuators (series elastic, Hill, DC motor) are integrated with the implicit midpoint rule, which conserves quadratic energy exactly. For the Hill model each step needs the root of a nonlinear equation; on each branch of the force-velocity curve it is a quadratic, which we solve in closed form.

\textbf{Reflex test.} A torque of 3.0~µN\,mm was applied to the femur-tibia joint of a leg for 200~ms, toward flexion or extension. For each case we averaged the late movement over replicates with motors on and with motors off, and took their ratio; the held ratio is the median of these ratios over cases. The actuator study used eight leg and direction cases with three replicates (ten for settings near the threshold); the null-network tests used three legs and both directions with four replicates. The reflex survives if the median ratio is below 0.9 and at least half the cases are reduced by more than 10\% [P18].

\subsection*{Walking stand-in}
NeuroMechFly v2 in FlyGym 2.1.1~\cite{wangchen2024} provided the walking body; the MaleCNS network, from a male, drove this female body [B8]. Because the connectome did not walk, legs were driven by the hybrid turning controller of NeuroMechFly v2 (Apache-2.0) [L2]. Descending neuron rates, averaged over 100~ms, set its two drive signals:
\begin{eqnarray}
D=\mathrm{clip}\!\left(\frac{r_{\mathrm{DNg100}}}{150\,\mathrm{Hz}},0,1\right),\quad
a=\frac{r_{\mathrm{DNa02},L}-r_{\mathrm{DNa02},R}}{50\,\mathrm{Hz}},\quad
u_{L,R}=D\left[1-0.8\,\mathrm{clip}(\pm a,0,1)\right].
\end{eqnarray}
When MDN fires above 20~Hz and above DNg100, both drives become $u=-\mathrm{clip}(r_{\mathrm{MDN}}/150\,\mathrm{Hz},0,1)$ and the fly walks backward. A tonic 200~Hz Poisson drive to DNg100 started walking. In the actuator study the controller's joint targets became torque commands, $\tau=\mathrm{clip}(45(q^*-q),\pm65)$, executed by each actuator [MX-2]; the leg motor neurons are not used in this mode.

\subsection*{Escape}
A labelled detector [E1] reads 10~ms bins of per-side rates. It signals take-off when the giant fibre on one side fires at 150~Hz or more (two spikes in a bin), or when the summed rates of DNp02 and DNp11 on one side reach 40~Hz. The two-spike rule was chosen after the first controls, in which a single spike gave false take-offs, and was then used online in the integrated fly [M8-1]. After take-off the walking command is set to zero; the walking body cannot fly.

\subsection*{Flight}
\textbf{Body.} The flight body is the model that flybody's flight-imitation task compiles~\cite{vaxenburg2025}, exported once with its fluid model, in centimetres, grams and seconds [FL1]; physics runs at 0.05~ms and control at 0.2~ms, with two network steps per control step. We reproduced flybody's published flights (100 of 100 reference episodes) before changing anything.
\textbf{Power muscle} [FL3]. The dorsal longitudinal and dorsoventral muscles form an antagonistic, stretch-activated pair on the wing stroke angle $\phi$:
\begin{eqnarray}
\tau_{sa}\dot s_i=-s_i+s_{\max}\tanh\!\left(\frac{\max(0,e_i)}{s_{\max}}\right), \quad e_{\mathrm{DLM}}=\phi-\phi_0=-e_{\mathrm{DVM}},
\end{eqnarray}
\begin{eqnarray}
\tau_c\dot c=-c+k_c r_{\mathrm{MN}}, \qquad \tau_{\mathrm{act}}=T_{sa}\,c_{\max}\tanh(c/c_{\max})\,(s_{\mathrm{DVM}}-s_{\mathrm{DLM}}),
\end{eqnarray}
with an elastic thorax $H_e=\tfrac12K(\phi-\phi_0)^2$ tuned to 218~Hz. Motor neuron spikes act only through slow calcium $c$ ($\tau_c=100$~ms, $k_c=0.02$ per Hz); stretch activation $s$ delays force by $\tau_{sa}=1.5$~ms, with $s_{\max}=0.5$. The gain $T_{sa}$ is $G=3$ times the gain at which stretch activation would just cancel the passive damping of hinge and air at a motor neuron rate of 20~Hz. The calcium ceiling $c_{\max}=0.3$ was used from the throttle study on; earlier runs had none. The thorax spring is the only storage; stretch activation and calcium are signal states that store no energy. Active power enters through a modulated source port and is booked in the ledger, which the muscle integrates with exponential steps. Unlike the leg actuators, this muscle is not passive by design: the delayed stretch activation acts as negative damping, which is what sustains the wingbeat. Muscle mass is 30\% of body mass~\cite{lehmann1997}. Wing rotation and deviation follow the hover recording through stiff servos.
\textbf{Steering} [FL6, FL7, FL8]. The wing hinge network of Melis et al.~\cite{melis2024}, ported to NumPy (largest difference from the original 1.0e-6), maps the activity of 12 steering muscles to wing kinematics; deviation and rotation are added to the servo targets and stroke acts as a per-wing gain. Eleven muscles are driven and one is held at baseline [FL8]. Each muscle's activity is $0.5+0.5\,\Delta r/60\,\mathrm{Hz}$, clipped to $[-0.5,1.5]$ and filtered over 50~ms, with $\Delta r$ the change in its motor neuron rate from rest [FL7].
\textbf{Halteres} [FL10, FL11]. Each haltere beats in antiphase with the wings; body rotation adds a Coriolis shear that shifts the spike phase of three afferent groups. In the stability tests a labelled torque $+c\,\omega$ about yaw and roll cancels part of the model's passive damping, lengthening its half-life to two wingbeats [FL14]; a labelled phase path lets module C afferents advance b1 [FL15].
\textbf{Flight stand-in} [FL16]. flybody's trained flight policy~\cite{vaxenburg2025} flew the body. Every 10~ms the connectome rewrote the reference path ahead of the fly. The left-right difference of DNa04 set turning rate (0.05~rad/s per Hz). Climb rate was set by the mean power motor neuron rate relative to its hover value (5~cm/s per unit, clipped to $\pm2$~cm/s, filtered over 50~ms); only the turning command comes from a descending neuron.

\subsection*{A published closed loop, rebuilt}
We rebuilt the closed loop of Minnick et al.~\cite{minnick2026} on BANC from their description. Six leg motor pools set the amplitude of each leg's walking oscillator; on top, a hand-written odour rule slows the legs on the side with more odour by a factor $1-0.8|e|v$, where $v$ is the MBON valence (approach minus avoidance MBON spikes over a normaliser, clipped at $\pm1.5$) and $e$ the left-right odour difference, and a looming rule turns away from walls. For BANC we used the paper's fixed normaliser of 37.35. For MaleCNS, which the paper did not run, the normaliser was the mean MBON difference at rest of each network, intact or null [C2-8]; this choice lets a scrambled network keep a valence near 1. Choices the paper leaves open, such as arena layouts, are in the register [C2-1 to C2-9]. Each fly ran nine arenas with 32 food sources for 20~s each; a source counted as reached within 2.5~mm. We ran 20 seeds per condition (10 for R2 and W) in four arms. Arm A was the loop as described, with the intact network, R0, a lesion of Kenyon cells and MBONs, and an open-loop replay of saved commands from a start rotated by 30 degrees. Arm A0 was the loop with the odour rule off. Arm Av was arm A with $v$ fixed at $+1$; it used the same random draws as arm A, so its comparison with A is paired by seed. Arm B was our walking stand-in in the same arenas, commanded by the descending neurons with no hand-written rule, with its own replay control. Kinematic controls were a random walk and a reference walker at the loop's median speed. Tests and the fidelity criterion (within 15 percentage points of the paper) were fixed before the runs; the valence analysis was found afterwards and then tested directly [C2-10].

\subsection*{Null networks and controls}
Every null family keeps each connection's source and synapse count, and so each neuron's sign and total output. R0 permutes targets across the nervous system. R1 permutes targets within a class group (class, intrinsic group and body side). R2 permutes within cell type, falling back to R1 for untyped neurons (25.5\% in BANC, 1.3\% in MaleCNS). R2seg also keeps the leg segment; it was applied to the leg network of the rhythm test. W keeps the graph and permutes synapse counts among each neuron's outputs. Duplicate targets created by a permutation are merged, which removes 0.24\% of connections under R0 and up to 6.3\% under R2, so in- and out-degrees are kept only approximately. We built five networks per family, and ten for R2seg and the reflex follow-up, with four seeds per network (eight for the drum follow-up). A result depends on the wiring at a level if the real network lies outside the range of every null network at that level. The test is two-sided, so with five networks this happens by chance one time in three, and with ten about one time in five; we therefore read single-level results with five networks as indications. Readouts were fixed in configuration files before any null network ran. Further controls were a sham actuator swap, the stand-in alone without the network, and silencing all nerve cord interneurons.

\subsection*{Statistics}
Two-sided Mann-Whitney U tests compared conditions (for the rebuilt loop, with Cliff's delta and a 95\% bootstrap interval), with Benjamini-Hochberg correction where many neuron types were tested; drum steering used a one-sided test in the predicted direction. One-sided Fisher's exact tests compared take-off counts. For halteres, $t$ is a regression slope over its standard error. We report medians or means with standard deviations across seeds and give the number of seeds with each result. Choices made after seeing data, such as the two-spike escape threshold and the held-ratio metric, are marked as such.

\subsection*{Software and compute}
Python 3.12 with NumPy 2.5.4, SciPy 1.18.1, pandas 3.1.1, MuJoCo 3.16.0~\cite{todorov2012} and FlyGym 2.1.1; flyvis 1.2.0 with PyTorch 2.15 on CPU; upstream flybody in Python 3.10 with TensorFlow 2.8. Batches ran on the H\'abr\'ok cluster of the University of Groningen (AMD EPYC nodes, no GPU), about 1,700 core-hours in total. Every run is set by a configuration file with explicit seeds, and its outputs record the git commit. Each figure is built by a script from summary files in the repository, and writes a record of its source files (with checksums) and of every number it plots.

\subsection*{Use of AI tools}
This study was carried out with Claude Code (Anthropic; model Claude Opus 5.5), an AI coding assistant, under the author's direction. It wrote most of the simulation and analysis code, ran simulations on the H\'abr\'ok cluster, built the figures from scripts, and drafted text that the author revised. The author set the research questions, approved every milestone and design decision, and reviewed all results. Outputs were checked by unit and regression tests, by tracing every reported number to a configuration, seed, git commit and checksummed data file, and by an assumption register (S1 Table). No data were generated by the AI tool other than through the simulations described in Materials and methods.

\section*{Supporting information}
\paragraph*{S1 Video.}
\label{S1_Video}
\textbf{The closed loop in action.} Re-rendered conditions from each part of the study, with captions giving the measured result.

\paragraph*{S1 Table.}
\label{S1_Table}
\textbf{Assumption register.} Every assumption, stand-in and labelled change, with its source and how it could be tested.

\section*{Acknowledgments}
We thank the BANC and MaleCNS (FlyEM) teams for releasing their connectomes under CC BY, and the authors of FlyGym, flybody, flyvis and the wing hinge network for releasing their code. We thank the Center for Information Technology of the University of Groningen for access to the H\'abr\'ok cluster.

\section*{Funding}
The author received no specific funding for this work.

\section*{Competing interests}
The author has declared that no competing interests exist.

\section*{Author contributions}
Mauricio Mu\~noz Arias: conceptualization, methodology, software, validation, formal analysis, investigation, data curation, writing (original draft, review and editing), visualization, project administration.

\bibliography{refs}

\end{document}